\documentclass[11pt]{article}
\usepackage{amsmath} 
\usepackage{amssymb}	
\usepackage{graphicx} 
\usepackage{multicol} 
\usepackage{pstricks}
\usepackage{authblk}
\usepackage{graphicx}
\usepackage[small]{caption}
\usepackage{subcaption}
\usepackage{float}
\usepackage{amssymb}
\usepackage{bbold}
\usepackage{color}
\usepackage{bbm}
\usepackage{fullpage}
\usepackage{amsmath}
\usepackage{amsthm}
\usepackage{appendix}
\usepackage[normalem]{ulem}
\usepackage{hyperref}
\hypersetup{
    colorlinks,
    citecolor=red,
    filecolor=blue,
    linkcolor=blue,
    urlcolor=blue
}

\newcommand{\tr}{\mbox{Tr}\,}
\providecommand{\norm}[1]{\lVert#1\rVert}

\newcommand{\inte}[4]{\int_{#1}^{#2} \! #4 \, \mathrm{d}#3}
\renewcommand{\d}[2]{\frac{d #1}{d #2}} 
\newcommand{\pd}[2]{\frac{\partial #1}{\partial #2}} 

\newcommand{\var}[2]{\frac{\delta #1}{\delta #2}} 

\title{Exploring the complexity of quantum control optimization trajectories}
\author[1]{Arun Nanduri}
\author[2]{Ofer M. Shir}
\author[1]{Ashley Donovan}
\author[1]{Tak-San Ho}
\author[1]{Herschel Rabitz}
\affil[1]{\emph{Department of Chemistry, Princeton University, Princeton, NJ 08544}}
\affil[2]{\emph{School of Computer Science, Tel-Hai College, Upper Galilee, Israel}}
\begin{document}
\maketitle

\begin{abstract}
  The control of quantum system dynamics is generally performed by seeking a suitable applied field. The physical objective as a functional of the field forms the quantum control landscape, whose \textit{topology}, under certain conditions, has been shown to contain no critical point suboptimal traps, thereby enabling effective searches for fields that give the global maximum of the objective. This paper addresses the \textit{structure} of the landscape as a complement to topological critical point features. Recent work showed that landscape structure is highly favorable for optimization of state-to-state transition probabilities, in that gradient-based control trajectories to the global maximum value are nearly straight paths. The landscape structure is codified in the metric $R\geq 1.0$, defined as the ratio of the length of the control trajectory to the Euclidean distance between the initial and optimal controls. A value of $R=1$ would indicate an exactly straight trajectory to the optimal observable value. This paper extends the state-to-state transition probability results to the quantum ensemble and unitary transformation control landscapes. Again, nearly straight trajectories predominate, and we demonstrate that $R$ can take values approaching 1.0 with high precision. However, the interplay of optimization trajectories with critical saddle submanifolds is found to influence landscape structure. A fundamental relationship necessary for perfectly straight gradient-based control trajectories is derived, wherein the gradient on the quantum control landscape must be an eigenfunction of the Hessian. This relation is an indicator of landscape structure and may provide a means to identify physical conditions when control trajectories can achieve perfect linearity. The collective favorable landscape topology and structure provide a foundation to understand why optimal quantum control can be readily achieved.
\end{abstract}

\section{Introduction}
\begin{figure}[h!]
  \centering
  \includegraphics[width=\textwidth]{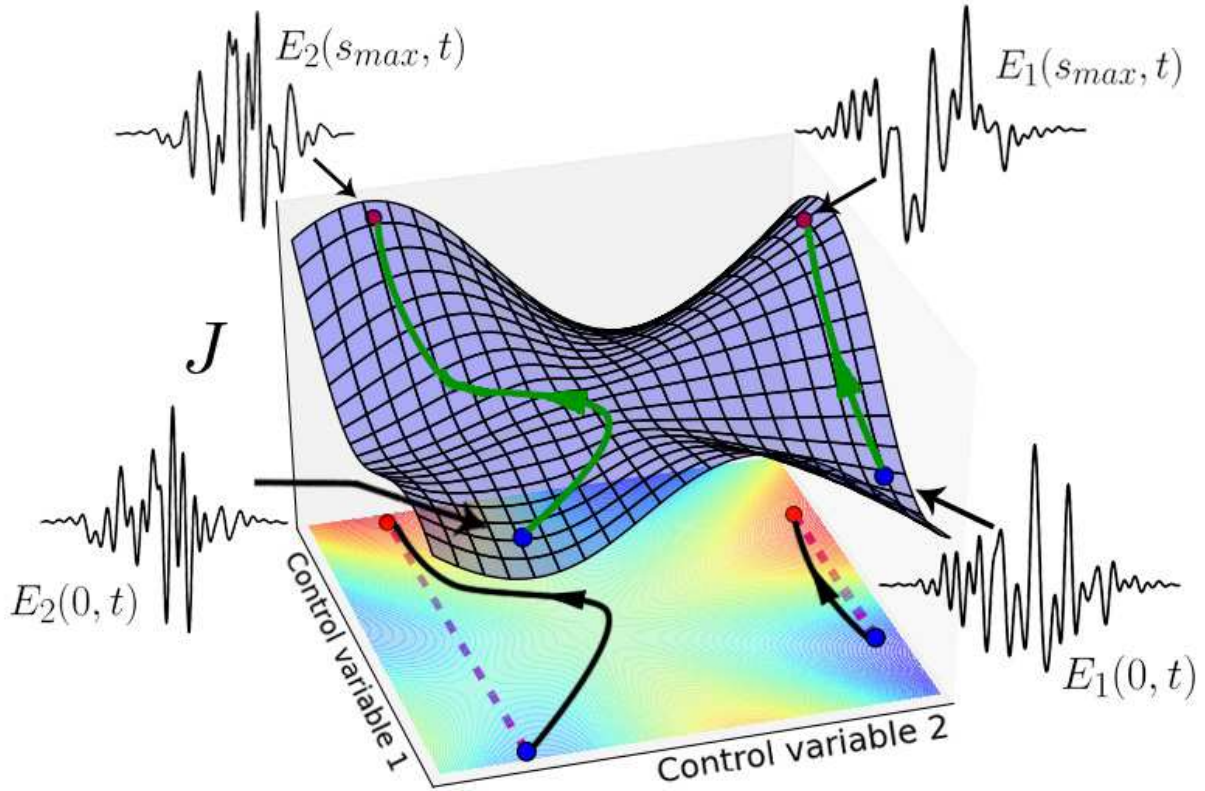}
  \caption{Illustrative sketch of a quantum control landscape (blue surface) and its underlying control space (colored contour map), also displayed as the projection onto the plane spanned by two control variables (in practice many more are used). The vertical axis represents a cost functional $J$ subject to maximization. The two optimization paths shown, in green, differ in the nature of their $R$ values. The path on the right, proceeding from $E_1(0,t)$ to $E_1(s_{max},t)$, climbs in a direct manner to the top of the control landscape. Consequently, its projection into control space, in black, has a path length very close to the Euclidean distance along the straight line between the projection's endpoints, shown as a magenta dashed line. Thus, this path's $R$ value is low. On the other hand, the optimization path on the left, from $E_2(0,t)$ to $E_2(s_{max},t)$, is attracted away from proceeding directly to a maximum by the saddle point at the center of the landscape. As a result, its projection into control space is much longer than the straight line between its endpoints, and so this path possesses a higher $R$ value. The paths do not strictly terminate at the top and bottom of the landscape, as optimizations generally begin and end at values of the cost functional which approximate $J^{min}$ and $J^{max}$, respectively, to acceptable precision.}
  \label{sketch}
\end{figure}

\indent Quantum optimal control theory (OCT) provides a basis to explore applied field tuning of quantum system dynamics~\cite{balint-kurti,HDR,Dong,Moorewithcontrolstuff,W}. Recent successes include achieving control of electrons in quantum dots through time-varying gate voltages~\cite{Blasi}, controlling electronic correlations in molecules~\cite{Nest}, manipulating the dynamics of quantum many-body systems~\cite{Doria}, cooling ultracold atomic gases~\cite{Rahmani}, eliciting quantum revivals~\cite{Rasanen}, and generating adiabatic time evolution~\cite{Young}. Optimal control experiments (OCE) have found growing success even for complex systems, especially through the use of closed-loop learning algorithms~\cite{lasers} combined with femtosecond laser pulse shaping technology~\cite{pulseshaping}. For example, these advances have enabled vibrational control of population inversion in Bose-Einstein condensates~\cite{Bucker}, strong-field ionization of silver atoms~\cite{Truong}, and coherent energy transfer in light-harvesting complexes~\cite{Hildner}. Underlying these achievements is the question of why desirable quantum controls can evidently be found with only modest search effort over the vast space of possible control fields. An important factor in addressing this question is the favorable control landscape \emph{topology}, which results upon satisfaction of certain physical assumptions~\cite{surveypaper}. The present paper considers a second contributing factor based on the landscape \emph{structure} (i.e., non-topological features) reflected in the behavior of the evolving controls during the optimization process. \\
\indent The dynamics of a closed quantum system interacting with an applied field is described by the time-dependent Schr{\"o}dinger equation
\begin{equation}\label{TDSE}
i \hbar \frac{\partial U(t,0)}{\partial t} = H(t)\, U(t,0), \phantom{space} U(0,0)=1,
\end{equation}
where $H(t)$ is the time-dependent Hamiltonian and $U(t,0)$ is the corresponding evolution operator (propagator) with $0\leq t \leq T$. Here $T$ is the time at which the objective is optimized. Treating the interaction within the dipole approximation, the time-dependent Hamiltonian is given by
\begin{equation}\label{Ham_dipole}
H(t) = H_0 - \mu E(t),
\end{equation}
where $H_0$ is the field-free Hamiltonian, $\mu$ is the dipole moment operator and $E(t)$ is the time-dependent control field. \\
\indent Through Eq.~(\ref{TDSE}), a control field can be mapped to a value of a cost functional $J[E(t)]$, which may be either the expectation value $J_O$ of an observable $O$ or the cost functional $J_W$ for creating a target unitary transformation $W$ (see Eqs.~(\ref{TrpO}) and~(\ref{Wcost}), respectively). The functional relationship between $J[E(t)]$ and the control field $E(t)$ can be cast in terms of a \textit{quantum control landscape}, where each point $J[E(t)]$ on the landscape corresponds to a particular control field $E(t)$, $0\leq t \leq T$. Finding successive control fields that eventually lead to an optimal value of the cost functional entails climbing the landscape. We will refer to `climbing' as moving towards the target value of $J$, regardless of whether the cost functional is to be maximized or minimized. Using a local search algorithm to climb (e.g., the gradient algorithm utilized in this work) results in smooth paths up the landscape $J[E(s,t)]$ that can be parametrized by the continuous variable $s\geq 0$. Every such path up the landscape is associated with a trajectory $E(s,t)$ through \textit{control space}. Thus, starting with an initial control field $E(0,t)$ that produces a low value of the cost functional, the goal is to find the final control field $E(s_{max},t)$ that yields a high (optimal) value of the cost functional at $s=s_{max}$. \\
\indent Many recent studies have considered the \textit{topology} of quantum control landscapes~\cite{surveypaper,Wsaddlepoints,Wsaddlepoints2,Wstructure,BigTrpO}. The control landscape topology is specified by the character of the critical points where $\frac{\delta J}{\delta E(t)}=0$, $\forall t\in [0,T]$. A critical point at a suboptimal value of $J$ where the Hessian $\frac{\delta^2 J}{\delta E(t) \delta E(t')}$ is negative semi-definite would indicate a (second order) trap capable of halting an effort attempting to reach the top of the landscape. The landscape topology has been analyzed for closed quantum systems with $N$ states under the assumptions of: (i) controllability, such that any unitary evolution matrix $U(T,0)$ can be generated by at least one control field at some sufficiently large time $T$, (ii) surjectivity, whereby the set of functions $\frac{\delta U_{ij}(T,0)}{\delta E(t)}$, $\forall i,j$ are linearly independent over $t\in [0,T]$, and (iii) full accessibility, allowing for free access to any control field. Upon satisfaction of these assumptions, the quantum control landscape can be shown to contain only trap-free intermediate value critical saddle submanifolds (when they exist for a particular application)~\cite{surveypaper,Wsaddlepoints,BigTrpO} along with the absolute minimum and absolute maximum submanifolds. Thus, the landscape topology is very favorable to performing OCEs or OCT simulations, which is an important factor in explaining the evident ease of locating optimal control fields. \\
\indent Beyond issues of topology, the \textit{structure} of the quantum control landscape away from the critical points could greatly influence the nature of a control trajectory winding its way to an optimal field. If landscape structure is highly gnarled, this circumstance could call for complex control trajectories, although the trap free topology assures that the path (e.g., guided by a gradient ascent algorithm) will lead to an optimal solution. The encountered landscape structure upon climbing may be examined by quantifying the linearity of the control trajectories~\cite{Nanduri}. A predilection towards nearly straight trajectories is indicative of very simple control landscape structural features. Prior experimental work aiming to maximize second harmonic generation found that all of the sampled control trajectories were strikingly close to being straight~\cite{roslund}. Recent theoretical work has also supported the presence of simple landscape structure by examining the linearity of control trajectories in simulations over the state-to-state transition probability landscape~\cite{Nanduri,complexity,hamiltonianstructure}.  \\
\indent In this work, we extend the state-to-state transition probability quantum control landscape structure analysis to consider two additional landscapes:
\begin{equation}\label{TrpO}
J_O = \tr \left (\rho(T) O \right ) = \tr \left ( U(T,0) \rho(0) U^{\dagger}(T,0) O \right ),
\end{equation}
and
\begin{equation}\label{Wcost}
J_W=\norm{W-U(T,0)}^2=2N - 2\mbox{Re} \bigg \{  \tr \left ( W^{\dagger}U(T,0) \right )  \bigg \}
\end{equation}
where $\norm{\cdot}$ denotes the Frobenius norm. The quantum ensemble control landscape $J_O$ allows for controlling arbitrary observables and consideration of systems which are initially in mixed states $\rho(0)$, and the unitary transformation landscape $J_W$ is relevant to the prospect of creating logic gates for quantum information processing~\cite{Nielsen}. Both landscapes can contain critical saddle submanifolds~\cite{Wsaddlepoints,BigTrpO}, which are not present in the simpler state-to-state transition probability landscape. Thus, we aim to assess the influence of these saddle submanifolds on landscape structure. Specifically, we will show that encountering saddle submanifolds at intermediate heights on the landscape can drive control trajectories away from straight paths. This phenomenon is sketched in Figure \ref{sketch}. The trajectory on the right takes a rather straight path up the landscape, while the trajectory on the left is initially attracted to the saddle point, thus altering its direct nature. Nevertheless, we will show in numerical simulations that the trajectories leading to optimal controls are generally still close to being straight, despite the presence of saddles. Lastly, we will derive a mathematical criterion showing that the gradient of the cost functional must be an eigenfunction of the Hessian in order to achieve a perfectly straight gradient-based control trajectory. \\
\indent The remainder of the paper is organized as follows. The linearity measure $R$ of a control trajectory is defined in Section 2. Section 3 gives the gradient-based search algorithm used in this work to identify control fields that optimize arbitrary observables or achieve unitary transformations, respectively, through $J_O$ or $J_W$. The presence of landscape saddle submanifolds is examined, and a metric is outlined that gives the distance from any point on a trajectory to a particular saddle submanifold. In Section 4, a derivation is presented specifying the Hessian-gradient eigen-relation for straight control trajectories. Section 5 provides numerical results on the statistical behavior of $R$ over a large ensemble of landscape trajectories for $J_O$ and $J_W$. We also illustrate satisfaction of the Hessian-gradient eigen-relation for nearly straight control trajectories. Concluding remarks are furnished in Section 6.

\section{The control trajectory linearity measure $R$}
\indent The goal of the present work is to explore quantum control landscape structure reflected in the nature of the control trajectories followed during optimizations. Figure~\ref{sketch} indicates that a landscape trajectory has a projected image in the underlying control space. As we use a myopic gradient algorithm to climb the landscape, the degree of gnarled character encountered on the climb is manifested in the trajectory through control space. To this end, we construct a ratio $R$ delineating the linearity of the control trajectory. Specifically, $R$ is the ratio of the path length of the control trajectory between the initial and final fields, given by
\begin{equation}\label{pathlength}
d_{PL} = \int_0^{s_{max}} \left [ \frac{1}{T} \int_0^T \left ( \frac{\partial E(s,t)}{\partial s} \right )^2 dt\right ]^{\frac{1}{2}} \; ds,
\end{equation}
to the Euclidean distance between the initial and final fields in control space, given by
\begin{equation}\label{Euclength}
d_{EL} = \left [ \frac{1}{T} \int_0^T \left [ E(s_{max},t) - E(0,t) \right ]^2 dt \right ]^{\frac{1}{2}},
\end{equation}
where $s$ parametrizes the control trajectory. $R$ is expressed as
\begin{equation}\label{ratio}
R = \frac{d_{PL}}{d_{EL}}=\frac{\int_0^{s_{max}} \left [ \int_0^T \left ( \frac{\partial E(s,t)}{\partial s} \right )^2 dt\right ]^{\frac{1}{2}} \; ds}{\left [ \int_0^T \left [ E(s_{max},t) - E(0,t) \right ]^2 dt \right ]^{\frac{1}{2}}}.
\end{equation}
The lower bound for the ratio is $R=1$, which holds for a trajectory that takes a straight path between $E(0,t)$ and $E(s_{max},t)$. A trajectory with $R=1$ corresponds to the simplest possible path a gradient optimization can take to climb the quantum control landscape $J[E(t)]$, and trajectories approaching this limit indicate a lack of encounters with complex structural features. High values of $R$ mean that a control trajectory meanders through control space along a curve whose total length is much greater than the distance between its two endpoints, signifying that the optimization path encountered rugged structural features on the landscape. \\
\indent In a recent work, we found values of $R-1$ as low as $\sim 10^{-4}$ for a five level state-to-state transition probability landscape~\cite{Nanduri}. By examining the distances between the endpoints of the control trajectories, we also observed that trajectories with low and high $R$ values are distributed almost identically throughout control space. Furthermore, we found that it was possible to characterize the complexity of low $R$ trajectories through a `straight shot' climbing procedure. This procedure operates by calculating the gradient of the objective (e.g., the state-to-state transition probability) at the initial control field, and then proceeds to evolve the field along that specified direction until a local maximum is encountered. For suitably low $R$ trajectories, nearly optimal control fields may be found in this way by merely continuing to march in the initially identified direction. This technique is useful for characterizing the straight nature of a trajectory, but it is still short of a constructive control algorithm without knowledge of a suitable initial field. \\ 

\section{Negotiating the quantum control landscape}\label{negotiating}
The quantum systems considered here have $N$-level Hamiltonians $H$ in Eq.~(\ref{Ham_dipole}) consisting of the field-free Hamiltonian $H_0$ as an $N\times N$ diagonal matrix, and $\mu$ as the dipole moment given by an $N\times N$ real symmetric matrix. We will utilize the gradient algorithm to prescribe the path of steepest ascent, thereby forming a natural ``rule'' to climb the landscape and consequently to navigate the underlying control space. The D-MORPH procedure will be used to implement the gradient search trajectory~\cite{DMORPH,otherDMORPH}. The process of climbing the landscape may be confounded by the presence of saddle submanifolds located at certain regions of specific intermediate heights on the landscape. The cartoon in Fig.~\ref{sketch} reflects this issue, and one goal of this work is to examine the degree to which saddles distort control trajectories. \\
\indent We parameterize a trajectory by a continuous variable $s\geq 0$, such that the control field becomes $E(s,t)$. As $s$ is increased, the change in the cost functional can be written using the chain rule as
\begin{equation}\label{DMORPH}
\d{J}{s} = \inte{0}{T}{t}{\var{J}{E(s,t)}\pd{E(s,t)}{s}}.
\end{equation}
Maximization of $J$ stipulates that $\d{J}{s}\geq 0$, which is assured by setting
\begin{equation}\label{gradientrule}
\pd{E(s,t)}{s}=\var{J}{E(s,t)}.
\end{equation}
Likewise, when minimizing $J$, we require $\d{J}{s}\leq 0$, so in that case we set $\pd{E(s,t)}{s}=-\var{J}{E(s,t)}$. Minimization applies to $J_W$, and $J_O$ may be minimized or maximized depending on the nature of the objective operator $O$. In the present work, the control variables adjusted by the gradient algorithm are the values of the control field $E(s,t_i)$ at a set of discrete time points $t_i$, $i=1, 2, \dotsc$. In order to ensure that each simulation for a particular application traverses the same domain up (or down) the quantum control landscape, every trajectory begins at the same initial height on the landscape $J^I$ and ends at the same final height on the landscape $J^F$. This criterion is met by using the D-MORPH algorithm to first guide a climb down (up) the landscape to $J^I$, if an initial trial field yields a value of $J$ that is above (below) $J^I$. The resultant field at $J^I$ is taken as $E(0,t)$ for the purpose of determining $R$. The final height on the landscape $J^F$ is then achieved by $E(s=s_{max},t)$, identified from application of D-MORPH starting with the initial field $E(0,t)$. \\
\indent Calculation of the gradient $\var{J}{E(s,t)}$ requires utilization of the relevant cost functional, either $J_O$ or $J_W$. Sections 3.1 and 3.2 summarize these details, further include the characterization of the critical submanifolds of the landscapes, and introduce a method to determine the distance of a point on the landscape to any particular critical submanifold. In all of the simulations, the assumptions underlying the favorable landscape topology are taken as satisfied~\cite{surveypaper,Wsaddlepoints,BigTrpO}; consistent with these assumptions, no landscape trajectories prematurely terminated before reaching $J^F$ in our simulations.

\subsection{Quantum ensemble control landscape}
\indent The quantum ensemble control landscape $J_O[E(s,t)]$ for optimizing an arbitrary observable with a mixed initial state is given by Eq.~(\ref{TrpO}). The gradient of the objective may be written as~\cite{BigTrpO}
\begin{equation}\label{TrpOgrad}
\begin{aligned}
\frac{\delta J_O}{\delta E(t)} &= \tr \left\{ O\var{U(T,0)}{E(t)}\rho(0)U^{\dagger}(T,0) + OU(T,0)\rho(0)\var{U^{\dagger}(T,0)}{E(t)}\right\} \\
&= \tr \left\{ O\var{U(T,0)}{E(t)}\rho(0)U^{\dagger}(T,0) - OU(T,0)\rho(0)U^{\dagger}(T,0)\var{U(T,0)}{E(t)}U^{\dagger}(T,0) \right\} \\
&= \tr \left\{ U^{\dagger}(T,0)[\rho(T),O]\var{U(T,0)}{E(t)}\right\},
\end{aligned}
\end{equation}
where we have made use of the relation $\frac{\delta}{\delta E(t)}\left(U^{\dagger}(T,0)U(T,0)\right) = \frac{\delta U^{\dagger}(T,0)}{\delta E(t)}U(T,0) + U^{\dagger}(T,0)\frac{\delta U(T,0)}{\delta E(t)} = 0$ in the second equality. Climbing the landscape is affected by utilization of Eq.~(\ref{TrpOgrad}) in Eq.~(\ref{gradientrule}). This process could possibly be hindered if the trajectory comes near a saddle submanifold where $\var{J_O}{E(s,t)}$ is small at an intermediate, suboptimal value of $J_O$. A representative sketch is given in Figure~\ref{sketch}, where the optimization path on the left draws near a saddle submanifold, located at the center of the blue landscape surface. In order to assess the impact of encountering saddle manifolds upon the value of $R$ for a control trajectory, we first briefly summarize the nature of the critical submanifolds of the $J_O$ landscape. \\
\indent At a critical point, the first order variation satisfies $\frac{\delta J_O}{\delta E(t)}=0$. Thus, examining the last line of Eq.~(\ref{TrpOgrad}), we see that if $\var{U(T,0)}{E(t)}$ is of full rank (i.e., the surjectivity assumption (ii) from Section 1 is satisfied), we must have $[\rho(T),O]=0$ at a critical point~\cite{singularities}. From this relation, it is possible to obtain the form of the critical matrices $U_o$, and from that information, the distances $D$ to a particular critical submanifold from any point on the landscape; Appendix~\ref{sadpO} contains further details. The distance metric can be used to judge if a control trajectory has strayed close to the saddle submanifolds, and therefore $D$ allows us to assess whether proximity to these manifolds affects the $R$ value of a trajectory.


\subsection{Unitary transformation control landscape}
\indent In quantum information science, implementing basic logic gate operations necessitates the control of unitary transformations~\cite{Nielsen}. We therefore consider the optimization of the entire propagator $U(T,0)$, seeking to attain $W$, which leads to the unitary transformation cost functional given by Eq.~(\ref{Wcost}). $J_W$ has a minimum value of 0, corresponding to $U=W$, and a maximum of $4N$, corresponding to $U=-W$, where $N$ is the number of states in the quantum system. Thus, we would like to minimize $J_W$. \\
\indent The gradient for the unitary transformation landscape is~\cite{HDR}
\begin{equation}\label{Wgradient}
\begin{aligned}
\frac{\delta J_W}{\delta E(t)} &= -\tr \left\{ W^{\dagger}\var{U(T,0)}{E(t)} + \var{U^{\dagger}(T,0)}{E(t)}W\right\} \\
&=-\tr \left\{ W^{\dagger}\var{U(T,0)}{E(t)} - U^{\dagger}(T,0)\var{U(T,0)}{E(t)}U^{\dagger}(T,0)W\right\} \\
&= -\tr\left\{\left(W^{\dagger}U(T,0)-U^{\dagger}(T,0)W\right) U^{\dagger}(T,0)\frac{\delta U(T,0)}{\delta E(t)}\right\}.
\end{aligned}
\end{equation}
Optimizations may be carried out on the unitary transformation landscape by combining Eqs.~(\ref{gradientrule}) and (\ref{Wgradient}). Trajectories climbing the $J_W$ landscape may pass by saddle submanifolds. In order to assess the effect of these manifolds on $R$, we briefly review the nature of the critical points of the unitary transformation landscape. \\
\indent From the last line of Eq.~(\ref{Wgradient}), we see that at a critical point, where $\frac{\delta J_W}{\delta E(t)}=0$, if $\frac{\delta U(T,0)}{\delta E(t)}$ is of full rank, we must have $W^{\dagger}U(T,0)=U^{\dagger}(T,0)W$. This implies that $W^{\dagger}U_o$ is Hermitian. $W^{\dagger}U_o$ is also a unitary matrix, so at a critical point, the eigenvalues of $W^{\dagger}U_o$ must be either 1 or -1. As a result, the critical values of $J_W$ in Eq.~(\ref{Wcost}) are $0, 4, \dotsc, 4N$, implying that there exist N+1 critical submanifolds, each of which correspond to a different number of 1's and -1's as the eigenvalues of $W^{\dagger}U_o$~\cite{Wsaddlepoints,Wsaddlepoints2}. A discussion of the distance metric $D$ between any point on the landscape and a particular critical submanifold is contained in Appendix~\ref{sadW}. As explained in Section 3.1, evaluating $D$ along a landscape climb to reach $W$ can reveal the influence of encountering saddles on the path to an optimal control solution. \\


\section{`Straight shot' eigen-relation}
\indent Section 3 laid out how to climb the quantum ensemble and unitary transformation control landscapes using a gradient algorithm, and summarized the topology of the critical points of these landscapes. Here, we present a relation which must be satisfied for `straight shot' control trajectories, utilizing the gradient algorithm for \emph{any} quantum control landscape. \\
\indent Consider a control trajectory that is perfectly straight, i.e., $R=1$. Any such `straight shot' trajectory can be written~\cite{Nanduri}
\begin{equation}\label{straightshot}
E(s,t) = E(0,t) + \rho (s)\Delta E(t),
\end{equation}
where $0\leq \rho (s)\leq 1$ is a monotonically increasing function of $s$ as a result of the gradient algorithm assuring that $dJ/ds\geq 0$ (i.e., without loss of generality we seek to maximize $J$), and $\Delta E(t)=E(s_{max},t)-E(0,t)$ depends only on time. Utilizing Eq.~(\ref{gradientrule}), the gradient function along this trajectory is
\begin{equation}\label{straightshotderivative}
\var{J}{E(s,t)}= \rho'(s)\; \Delta E(t),
\end{equation}
where $\rho'(s)=\frac{d\rho(s)}{ds}$. \\
\indent According to Eq.~(\ref{gradientrule}), the instantaneous tangent of a control trajectory, $\pd{E(s,t)}{s}$, ``points'' in the same direction in control space as the gradient of the quantum control landscape, $\var{J}{E(s,t)}$, at location $s$. Furthermore, along a straight control trajectory, the gradient function must point in the \textit{same} direction at \textit{every} location on the trajectory; thus, the gradient functions $\var{J}{E(s,t)}$ and $\var{J}{E(s',t)}$ at two different locations $s$ and $s'$ on a straight control trajectory are related by a scale factor. Additionally, for straight trajectories, the difference between any two fields on the trajectory, $E(s,t)$ and $E(s',t)$ is always proportional to the same function $\Delta E(t)$. When using a gradient-based climbing algorithm, the function $\Delta E(t)$ is proportional to the gradient itself, as shown by Eq.~(\ref{straightshotderivative}). These facts allow us to conclude that an eigen-relation exists between the Hessian and the gradient on a straight control trajectory~\cite{eig1,eig2}. \\
\indent To make these considerations more specific, we expand the gradient of the cost functional $J[E(s+ds,t)]$ to first order in $ds$ along an arbitrary control trajectory at location $s$
\begin{equation}\label{expand}
\var{J}{E(s+ds,t)}=\var{J}{E(s,t)}+ds\inte{0}{T}{t'}{\frac{\delta^2 J}{\delta E(s,t') \delta E(s,t)} \pd{E(s,t')}{s}}.
\end{equation}
Here, $\frac{\delta^2 J}{\delta E(s,t') \delta E(s,t)}=\mathcal{H}(s;t,t')$ is the Hessian. Now assume that $R=1$ for the control trajectory $E(s,t)$. If we bring $\var{J}{E(s,t)}$ onto the left hand side, divide by $ds$, and take the limit, we have
\begin{equation}
\lim_{ds\to 0}\frac{1}{ds}\left(\var{J}{E(s+ds,t)}-\var{J}{E(s,t)}\right) = \d{}{s} \var{J}{E(s,t)} = \rho^{\prime\prime}(s)\Delta E(t).
\end{equation}
where Eq.~(\ref{straightshotderivative}) was used for the second equality. Using Eq.~(\ref{straightshotderivative}), we can rewrite the result as $\rho''(s)\Delta E(t) = \frac{\rho''(s)}{\rho'(s)}\var{J}{E(s,t)}$. On the right hand side of Eq.~(\ref{expand}), if we use Eq.~(\ref{gradientrule}), we obtain
\begin{equation}\label{eigenrelation0}
\frac{\rho''(s)}{\rho'(s)}\var{J}{E(s,t)} = \inte{0}{T}{t'}{\frac{\delta^2 J}{\delta E(s,t') \delta E(s,t)} \var{J}{E(s,t)}},
\end{equation}
or more compactly, with the time dependence implicitly understood,
\begin{equation}\label{eigenrelation}
\mathcal{H}(s)\cdot g(s)=\frac{\rho''(s)}{\rho'(s)}g(s),
\end{equation}
where $g(s)$ is the gradient of the cost functional, $\var{J}{E(s,t)}$, and the scalar product $\mathcal{H}(s)\cdot g(s)$ denotes the integral operation in Eq.~(\ref{eigenrelation0}). As the action of the Hessian on the gradient merely rescales the gradient, this equation states that the gradient is an eigenfunction of the Hessian along a straight control trajectory, with an associated eigenvalue $\frac{\rho''(s)}{\rho'(s)}$. We will illustrate this relationship in Section 5.3. In addition, one may show that repeating the above procedure by expanding the Hessian about the point $s$ and making use of Eq.~(\ref{straightshot}) yields a higher order eigen-relation, stating that the gradient is an eigenfunction of the third-order Hessian. Iterating this process yields a hierarchy of eigen-relations between the gradient and all higher-order derivatives of $J$; however, all of these relationships contain no more information than Eq.~(\ref{eigenrelation}), since the entire hierarchy holds upon satisfaction of Eqs.~(\ref{gradientrule}) and (\ref{straightshot}). \\
\indent Near a critical point, the eigen-relation is always true, even for trajectories with $R>1$. To see this relation, expand the gradient of $J[E(s_0+ds,t)]$ to first order about the location $s_0$, which is infinitesimally close to a critical point at $E(s_c,t)$
\begin{equation}
\var{J}{E(s_c,t)}\equiv \var{J}{E(s_0+ds,t)} = \var{J}{E(s_0,t)} + ds \inte{0}{T}{t'}{\frac{\delta^2 J}{\delta E(s_0,t') \delta E(s_0,t)} \pd{E(s_0,t')}{s_0}}
\end{equation}
where $ds=s_c-s_0$. Utilizing the gradient rule in Eq.~(\ref{gradientrule}) inside the integral and noting that $\var{J}{E(s_c,t)}=0$ at a critical point, we have
\begin{equation}\label{criticaleigenvector}
\var{J}{E(s_0,t)}= -ds\inte{0}{T}{t'}{\frac{\delta^2 J}{\delta E(s_0,t') \delta E(s_0,t)} \var{J}{E(s_0,t')}}.
\end{equation}
Equation~(\ref{criticaleigenvector}) shows that the gradient function is always an eigenfunction of the Hessian in the immediate neighborhood of a critical point, whether or not the global trajectory is a straight shot. When $R=1$, however, this behavior extends to the entire control trajectory. From the definition of $ds$, one can see that if $s_c<s_0$, so that the gradient trajectory is leaving the critical manifold, the eigenvalue in Eq.~(\ref{criticaleigenvector}) is positive, and likewise if $s_c>s_0$, so that the trajectory is arriving at the critical manifold, the eigenvalue is negative. This circumstance will be shown in Section 5.3 for the case of a trajectory moving away from the critical point at the bottom of the landscape where $ds<0$ and then transitioning to the neighborhood of the top of the landscape where $ds>0$. Thus, along a straight shot from the bottom to the top of the landscape, the Hessian eigenvalue associated with the gradient will change sign (i.e., at least once, and more often if saddle critical points are closely approached) at some point along the trajectory. In addition, the eigen-relation is nominally an independent consideration at each value of $s$. However, the diffeomorphic nature of the gradient climb over $s$ in Eq.~(\ref{gradientrule}) implies that a smooth eigen-relation should exist, regardless of which eigenvector (corresponding to the gradient function) initiates the climb. This behavior is affirmed in the simulations in Section 5.3.

\indent We note that the relations examined in this section were derived without making use of quantum mechanics, as the analysis leading to the gradient being an eigenfunction of the Hessian only utilizes the gradient climbing rule, Eq.~(\ref{gradientrule}), and the straight shot condition Eq.~(\ref{straightshot}). The interest in this paper is directed towards applications where $J[E(s,t)]$ arises in a quantum control context.

\section{Illustrations}
\indent This section provides numerical illustrations of the landscape structure analysis tools outlined in Sections 2-4. Section 5.1 shows that for randomly chosen control trajectories, the distribution of $R$ takes on very modest values, which did not exceed 2.0 in the present examples. The results also demonstrate that control trajectories with either very low or very high values of $R$ are spread in a like fashion throughout control space. Section 5.2 addresses the effect of saddle submanifolds upon optimization paths. Encountering saddle critical submanifolds on the way from the bottom to the top along an optimization path tends to modestly increase the associated $R$ value. Finally, Section 5.3 provides numerical evidence for the validity of the Hessian-gradient eigenrelation in Section 4. Dimensionless units are used throughout the simulations.

\subsection{Random sampling of control trajectories}
\indent In order to assess the structural complexity of the quantum ensemble and unitary transformation control landscapes, we recorded the $R$ values of landscape trajectories constructed using randomly chosen initial control fields. 

\subsubsection{Quantum ensemble control landscape}
\begin{figure}[h!]
  \centering
  \includegraphics[width=\textwidth]{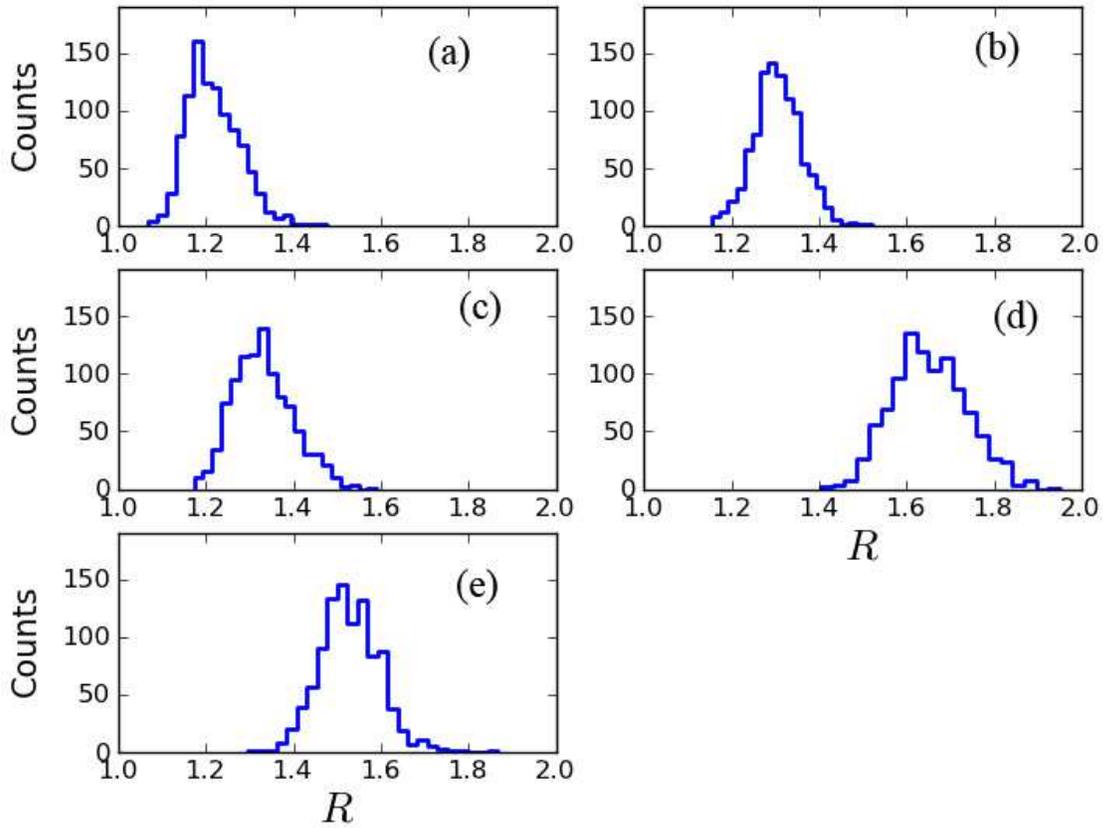}
  \caption{The distribution of $R$ values for five different quantum ensemble control landscapes. The cases are generated using: (a)~$\rho_1$ and $O_1$; (b)~$\rho_1$ and $O_2$; (c)~$\rho_2$ and $O_1$; (d)~$\rho_2$ and $O_2$; and (e)~$\rho_3$ and $O_3$. For each of these runs, the Hamiltonian $H_0$ and dipole $\mu$ of Eq.~(\ref{H0big}) were used. All of the distributions skew towards the right, showing that $R$ tends to accumulate near 1.0. 1000 optimizations were carried out for each landscape.}
  \label{ratios6}
\end{figure}

\begin{figure}[h!]
  \centering
  \includegraphics[width=\textwidth]{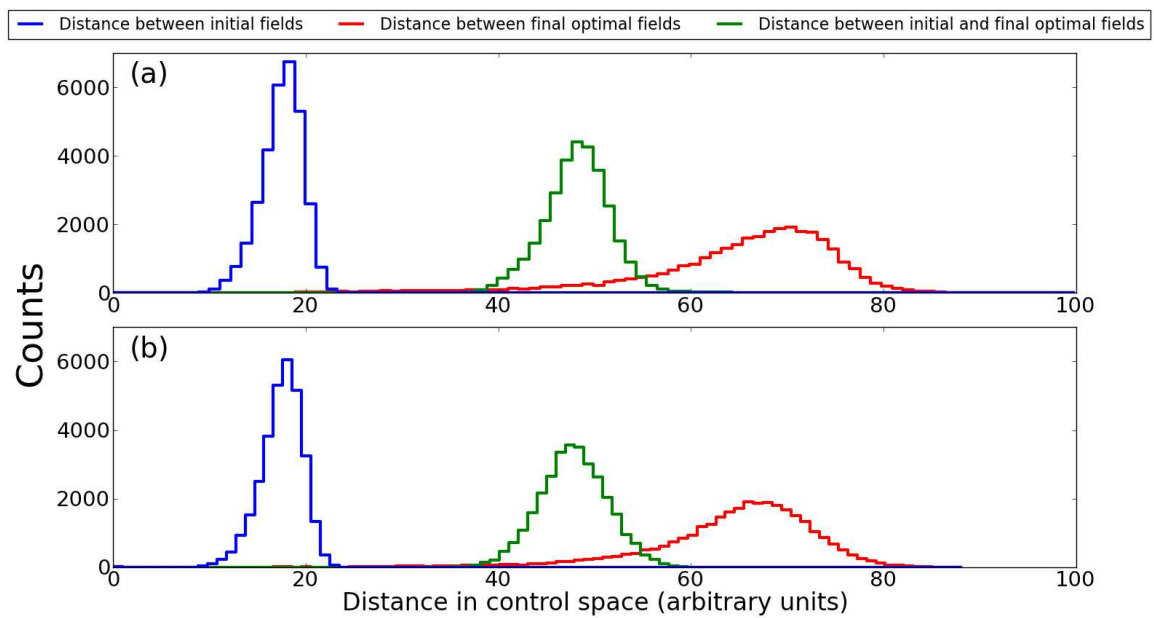}
  \caption{The distributions of pairwise distances between the initial fields, final fields, and initial-final fields for the landscape generated by $\rho_1$ and $O_1$, from Eqs.~(\ref{rhos12}) and (\ref{Os12}), respectively. The distances are calculated with Eq.~(\ref{pairwise}). Plot (a) was produced using the 250 optimizations with the lowest values of $R$, while plot (b) was produced using the 250 optimizations that had the highest values of $R$. The similarity between the two plots suggests that low $R$ and high $R$ trajectories are distributed in essentially the same way throughout control space.}
  \label{disthistpO}
\end{figure}

\indent Here we consider several eight-level quantum systems to illustrate the landscape structural findings. Five different quantum ensemble control landscapes are sampled, and 1000 optimizations of $J_O$ are performed over each landscape. Two different initial density matrices are used in generating these landscapes:
\begin{equation}
\rho_1=\begin{pmatrix} 1 & 0 & 0 & 0 & 0 & 0 & 0 & 0 \\ 0 & 0 & 0 & 0 & 0 & 0 & 0 & 0 \\ 0 & 0 & 0 & 0 & 0 & 0 & 0 & 0\\ 0& 0 & 0 & 0 & 0 & 0 & 0 & 0 \\ 0 & 0 & 0 & 0 & 0 & 0 & 0 & 0 \\ 0 & 0 & 0 & 0 & 0 & 0 & 0 & 0 \\ 0 & 0 & 0 & 0 & 0 & 0 & 0& 0 \\ 0 & 0 & 0 & 0 & 0 & 0 & 0 & 0
\end{pmatrix},
\phantom{space}
\rho_{2}=\begin{pmatrix} \frac{1}{4} & 0 & 0 & 0 & 0 & 0 & 0 & 0 \\ 0 & \frac{1}{4} & 0 & 0 & 0 & 0 & 0 & 0 \\ 0 & 0 & \frac{1}{4} & 0 & 0 & 0 & 0 & 0\\ 0 & 0 & 0 & \frac{1}{4} & 0 & 0 & 0 & 0 \\ 0 & 0 & 0 & 0 & 0 & 0 & 0 & 0 \\ 0 & 0 & 0 & 0 & 0 & 0 & 0 & 0 \\ 0 & 0 & 0 & 0 & 0 & 0 & 0 & 0 \\ 0 & 0 & 0 & 0 & 0 & 0 & 0 & 0
\end{pmatrix}.
\label{rhos12}
\end{equation}
As the nature of a landscape's saddle manifolds is determined by the number and degree of degeneracies of the initial density matrix and observable matrix~\cite{BigTrpO,Rebing,TrpOtopology}, we will consider combinations of matrices with varying degrees of degeneracy so as to examine a broad class of landscapes. Increasing numbers of nondegenerate eigenvalues generally correspond to larger numbers of saddle submanifolds. In particular, we are interested in the effects of such submanifolds upon landscape structure. The density matrix $\rho_1$, which corresponds to the case of a pure initial state, represents an almost completely degenerate density matrix. In contrast, $\rho_2$ contains two eigenvalues which are both four-fold degenerate. The observables used in conjunction with these density matrices are
\begin{equation}
O_{1}=\begin{pmatrix} 0 & 0 & 0 & 0 & 0 & 0 & 0 & 0 \\ 0 & 0 & 0 & 0 & 0 & 0 & 0 & 0 \\ 0 & 0 & 0 & 0 & 0 & 0 & 0 & 0\\ 0& 0 & 0 & 0 & 0 & 0 & 0 & 0 \\ 0 & 0 & 0 & 0 & 0 & 0 & 0 & 0 \\ 0 & 0 & 0 & 0 & 0 & 0 & 0 & 0 \\ 0 & 0 & 0 & 0 & 0 & 0 & \frac{4}{9} & 0 \\ 0 & 0 & 0 & 0 & 0 & 0 & 0 & \frac{5}{9}
\end{pmatrix},
\phantom{space}
O_{2}=\begin{pmatrix} 0 & 0 & 0 & 0 & 0 & 0 & 0 & 0 \\ 0 & 0 & 0 & 0 & 0 & 0 & 0 & 0 \\ 0 & 0 & 0 & 0 & 0 & 0 & 0 & 0\\ 0& 0 & 0 & 0 & 0 & 0 & 0 & 0 \\ 0 & 0 & 0 & 0 & \frac{4}{17} & 0 & 0 & 0 \\ 0 & 0 & 0 & 0 & 0 & \frac{4}{17} & 0 & 0 \\ 0 & 0 & 0 & 0 & 0 & 0 & \frac{4}{17} & 0 \\ 0 & 0 & 0 & 0 & 0 & 0 & 0 & \frac{5}{17}
\end{pmatrix}.
\label{Os12}
\end{equation}
As these observable matrices also exhibit varying levels of degeneracy, the pairs $\rho_1$ and $O_1$, $\rho_1$ and $O_2$, $\rho_2$ and $O_1$, $\rho_2$ and $O_2$ are used to generate four distinct landscapes. Lastly, we also utilized
\begin{equation}\label{ND}
\rho_3=\begin{pmatrix} \frac{7}{28} & 0 & 0 & 0 & 0 & 0 & 0 & 0 \\ 0 & \frac{6}{28} & 0 & 0 & 0 & 0 & 0 & 0 \\ 0 & 0 & \frac{5}{28} & 0 & 0 & 0 & 0 & 0\\ 0 & 0 & 0 & \frac{4}{28} & 0 & 0 & 0 & 0 \\ 0 & 0 & 0 & 0 & \frac{3}{28} & 0 & 0 & 0 \\ 0 & 0 & 0 & 0 & 0 & \frac{2}{28} & 0 & 0 \\ 0 & 0 & 0 & 0 & 0 & 0 & \frac{1}{28} & 0 \\ 0 & 0 & 0 & 0 & 0 & 0 & 0 & 0
\end{pmatrix}
\phantom{space}
O_3=\begin{pmatrix} 0 & 0 & 0 & 0 & 0 & 0 & 0 & 0 \\ 0 & \frac{1}{28} & 0 & 0 & 0 & 0 & 0 & 0 \\ 0 & 0 & \frac{2}{28} & 0 & 0 & 0 & 0 & 0\\ 0 & 0 & 0 & \frac{3}{28} & 0 & 0 & 0 & 0 \\ 0 & 0 & 0 & 0 & \frac{4}{28} & 0 & 0 & 0 \\ 0 & 0 & 0 & 0 & 0 & \frac{5}{28} & 0 & 0 \\ 0 & 0 & 0 & 0 & 0 & 0 & \frac{6}{28} & 0 \\ 0 & 0 & 0 & 0 & 0 & 0 & 0 & \frac{7}{28}
\end{pmatrix},
\end{equation}
as an example of matrices with fully nondegenerate spectra. The optimizations used the Hamiltonian and dipole operators
\begin{multline}\label{H0big}
H_0=\begin{pmatrix} -10 & 0 & 0 & 0 & 0 & 0 & 0 & 0 \\ 0 & -8 & 0 & 0 & 0 & 0 & 0 & 0 \\ 0 & 0 & -4 & 0 & 0 & 0 & 0 & 0\\ 0& 0 & 0 & 2 & 0 & 0 & 0 & 0 \\ 0 & 0 & 0 & 0 & 10 & 0 & 0 & 0 \\ 0 & 0 & 0 & 0 & 0 & 20 & 0 & 0 \\ 0 & 0 & 0 & 0 & 0 & 0 & 32 & 0 \\ 0 & 0 & 0 & 0 & 0 & 0 & 0 & 46
\end{pmatrix}
\\
\mu=\begin{pmatrix} 0 & \pm 1 & \pm 0.5 & \pm 0.5^2 & \pm 0.5^3 & \pm 0.5^4 & \pm 0.5^5 & \pm 0.5^6 \\ \pm 1 & 0 & \pm1 & \pm 0.5 & \pm 0.5^2 & \pm 0.5^3 & \pm 0.5^4 & \pm 0.5^5 \\ \pm 0.5 & \pm 1 & 0 & \pm 1 & \pm 0.5 & \pm 0.5^2 & \pm 0.5^3 & \pm 0.5^4\\ \pm 0.5^2 & \pm 0.5 & \pm 1 & 0 & \pm 1 & \pm 0.5 & \pm 0.5^2 & \pm 0.5^3 \\ \pm 0.5^3 & \pm 0.5^2 & \pm 0.5 & \pm 1 & 0 & \pm 1 & \pm 0.5 & \pm 0.5^2 \\ \pm 0.5^4 & \pm 0.5^3 & \pm 0.5^2 & \pm 0.5 & \pm 1 & 0 & \pm 1 & \pm 0.5 \\ \pm 0.5^5 & \pm 0.5^4 & \pm 0.5^3 & \pm 0.5^2 & \pm 0.5 & \pm 1 & 0 & \pm 1 \\ \pm 0.5^6 & \pm 0.5^5 & \pm 0.5^4 & \pm 0.5^3 & \pm 0.5^2 & \pm 0.5 & \pm 1 & 0
\end{pmatrix}
\end{multline}
where the signs on the dipole matrix elements were chosen randomly under the constraint of $\mu$ remaining symmetric. Once these signs were chosen, they remained fixed for the 1000 optimizations over each individual landscape. \\
\indent The initial fields of these optimizations were parametrized as
\begin{equation}\label{parametrized}
E(t)=\frac{1}{F} \mbox{exp}[-0.3 (t-\frac{T}{2})^2] \sum_{n=1}^{M} a_{n} \, \mbox{sin}(\omega_n t + \phi_{n}),
\end{equation}
with the target time $T$ being 10. The amplitudes $a_n$ and phases $\phi_n$ were chosen randomly from the uniform distributions $[0,1]$ and $[0,2\pi]$, respectively. The frequencies are $\omega_n=n$ and $M=60$ to ensure that the initial control field was nearly resonant with every possible transition in the Hamiltonian of Eq.~(\ref{H0big}). $F$ is a normalization factor, picked so that the initial fluence of each field is 1. This normalization was chosen to avoid operating with strong initial fields. After the initial choice of the control field, the time interval $[0,T]$ was discretized into 1001 evenly spaced time points, and the value of the field at each of these time points served as the control variables. We define the limits of the optimizations on the landscape as follows. We denote $J_O^{\mathrm{max}}$ as the maximum value that $J_O=\tr (\rho(T) O)$ can attain for a particular pair of $\rho (0)$ and $O$, and $J_O^{\mathrm{min}}$ denotes the minimum value. When a random field drawn from Eq.~(\ref{parametrized}) is used to calculate $J_O=\tr (\rho(T) O)$, if $J_O<J_O^I=J_O^{\mathrm{min}} + 0.01[J_O^{\mathrm{max}}-J_O^{\mathrm{min}}]$, then D-MORPH is used to climb until $J_O=J_O^I$. Likewise, if $J_O>J^I_O$, D-MORPH is used to descend the landscape until $J_O=J^I_O$. Then, the resultant control field is optimized using D-MORPH until $J_O=J_O^F=J_O^{\mathrm{max}}-0.01[J_O^{\mathrm{max}}-J_O^{\mathrm{min}}]$. This latter trajectory, which takes the field from a height of $J_O^I$ to a height of $J_O^F$ on the landscape, is used to calculate $R$. The large window from $J_O^I\to J_O^F$ is chosen so that each of the optimizations proceeds over a wide swathe of the landscape and thereby encounters as many structural features as possible. In this regard, we have numerically shown that for the state-to-state transition probability landscape, pushing the fidelity of $J$ closer to either 0 or 1 at the beginning and end of the optimizations, respectively, has a diminutive effect on $R$~\cite{Nanduri}. \\
\indent Histograms of the resulting $R$ values from 1000 optimizations over the five different landscapes are presented in Figure~\ref{ratios6} for the cases (a) $\rho_1$ and $O_1$, (b) $\rho_1$ and $O_2$, (c) $\rho_2$ and $O_1$, (d) $\rho_2$ and $O_2$, and (e) $\rho_3$ and $O_3$. The distributions are skewed towards the right, indicating that the $R$ values tend to accumulate near 1.0. The smallest values are found in the histogram of Figure~\ref{ratios6}(a) which has a mean of $R=1.22$, followed by Figures~\ref{ratios6}(b) and \ref{ratios6}(c), which have mean values of $R=1.31$ and $R=1.33$, respectively. The landscapes sampled in these Figures~\ref{ratios6}(a), (b), and (c) correspond to $\rho$ and $O$ matrices with high degrees of degeneracy. The highest values of $R$ are found in the histograms of Figs.~\ref{ratios6}(d) and \ref{ratios6}(e), which have mean values of $R=1.65$ and $R=1.53$, respectively. The landscape sampled in Fig.~\ref{ratios6}(d), generated from $\rho_2$ and $O_2$, corresponds to low degree of degeneracy, and the case in Fig.~\ref{ratios6}(e) with $\rho_3$ and $O_3$ is fully nondegenerate. The statistics indicate that as the degree of degeneracy of the eigenvalues of $\rho(0)$ and $O$ is reduced (i.e., the number of saddle submanifolds increases), the underlying quantum control landscapes become more structurally complex. However, the distinctions between the encountered structural features are modest over all of the cases examined. In an absolute sense, all of the trajectories produced $R<2.0$ indicating only very mild landscape structural features. Prior experience with state-to-state transition probability landscapes suggests that `straight shot' trajectories are reserved for cases with $R<1.05$, and a search is generally needed to find a proper $E(0,t)$ to initiate such a trajectory~\cite{Nanduri}. \\
\indent A natural question is whether control trajectories that possess low or high $R$ values are concentrated in some particular region(s) of control space. As the initial controls in the statistical sampling considered above were randomly chosen, this assessment directly bears on the investigation of the range of $R$ values encountered. To address this question, Euclidean distances were calculated between (1) all pairs of initial fields, (2) all pairs of final fields, and (3) all initial and final fields from the collection of control trajectories for the landscape associated with $\rho_1$ and $O_1$. These distances were calculated using
\begin{equation}\label{pairwise}
d_{EL}^{ij} = \left [ \frac{1}{T}\int_0^T [E_i(t)-E_j(t)]^2 dt \right ]^{\frac{1}{2}}
\end{equation}
for control fields $E_i$ and $E_j$, and the results are displayed in Figure \ref{disthistpO}. Importantly, in computing these distances, we first used only the 250 trajectories with the lowest values of $R$ out of the 1000 trajectories that were sampled in generating Figure~\ref{ratios6}(a). Then, separately, we used the 250 trajectories with the highest values of $R$. In particular, trajectories with $R<1.173$ were used to generate the histograms in Figure~\ref{disthistpO}(a), and likewise with $R>1.256$ for the histograms in Figure~\ref{disthistpO}(b); in both cases the histograms show all possible pairs of distances, with $250\times 249/2\sim 3.1\times 10^4$ pairs. The distribution of distances between initial fields, in blue, is centered to the left and narrower than the distribution corresponding to the distances between optimal fields, in red. This behavior implies that the optimal fields are scattered more widely throughout control space than the randomly picked initial fields. The histogram in green of the distances between all initial-final field pairs is centered between the other two distributions. The most significant feature in Figure~\ref{disthistpO} is that the distributions are very similar, whether generated from trajectories with low values of $R$ (Figure~\ref{disthistpO}(a)) or from trajectories with high values of $R$ (Figure~\ref{disthistpO}(b)). This result suggests that trajectories with low and high values of $R$ are distributed in the same way across control space. Thus, straighter, more direct trajectories and those less so are likely to be equidistant from a trajectory originating from a random initial control field.

\subsubsection{Unitary transformation control landscape}
\begin{figure}[h!]
  \centering
  \includegraphics[width=.6\textwidth]{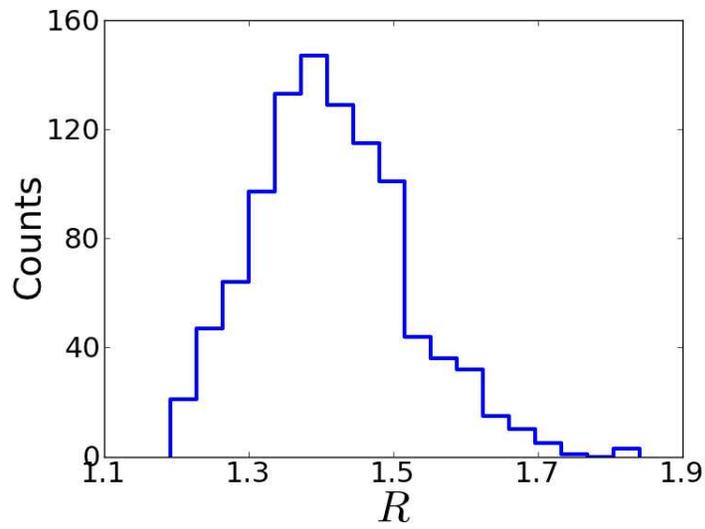}
  \caption{The distribution of $R$ values for a unitary transformation control landscape. 1000 optimizations were carried out, using the target unitary transformation $W$ of Eq.~(\ref{W}) and Hamiltonian $H_0$ and dipole $\mu$ matrices of Eq.~(\ref{H04}). Modest values of $R$, all less than 2.0, characterize this landscape's structure as simple.}
  \label{ratiosW}
\end{figure}

\begin{figure}[h!]
  \centering
  \includegraphics[width=\textwidth]{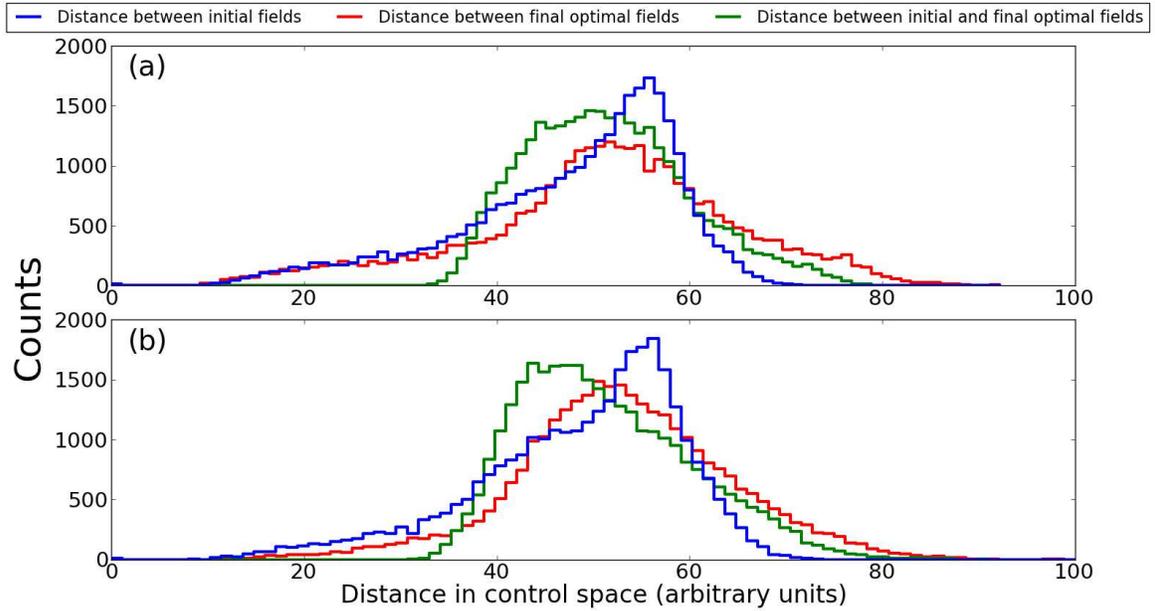}
  \caption{The distributions of pairwise distances between initial fields, final fields, and initial-final fields for the unitary transformation control landscape, calculated using Eq.~(\ref{pairwise}). Plot (a) was produced using the 250 optimizations with the lowest values of $R$, while plot (b) was produced using the 250 optimizations that had the highest values of $R$. The similarity between the sets of distributions in (a) and (b) suggests that low $R$ and high $R$ trajectories, as well as all of the associated fields, are distributed in the same way throughout control space.}
  \label{disthistW}
\end{figure}

\indent Here we consider a four-level quantum system for the target unitary transformation
\begin{equation}\label{W}
W=\begin{pmatrix} 1 & 0 & 0 & 0 \\ 0 & 0 & 1 & 0 \\ 0 & -1 & 0 & 0\\ 0 & 0 & 0 & 1
\end{pmatrix}
\end{equation}
with the Hamiltonian and dipole matrices being
\begin{equation}
H_0=
\begin{pmatrix} -10 & 0 & 0 & 0 \\ 0 & -7 & 0 & 0 \\ 0 & 0 & -1 & 0\\ 0 & 0 & 0 & 8
\end{pmatrix}
\phantom{space}
\mu=
\begin{pmatrix} 0 & \pm 1 & \pm 0.5 & \pm 0.5^2 \\ \pm 1 & 0 & \pm 1 & \pm 0.5 \\ \pm 0.5 & \pm 1 & 0 & \pm 1 \\ \pm 0.5^2 & \pm 0.5 & \pm 1 & 0
\end{pmatrix}
\label{H04}
\end{equation}
where the random signs on the dipole matrix were again chosen so that it remained symmetric. A total of 1000 initial fields were randomly chosen according to Eq.~(\ref{parametrized}), with $T=10$ and $M$ set to 20. The initial fields were adjusted as explained in Section~5.1.1 to arrive at a value of $J_W^I=0.99\times 16=15.84$, and the final achieved value after optimization was $J_W^F=0.01\times 16=0.16$. Thus, in this case we climb down the landscape as a minimization process. Figure~\ref{ratiosW} is a histogram of the $R$ values collected in this fashion. The plot is skewed right with a mean of $R=1.41$, indicating that the unitary transformation control landscape contains minimally complex structural features. The quantum ensemble landscapes considered earlier were for eight level systems, while this unitary transformation landscape is for a four level system. Together with previous work which observed that typical $R$ values for generating unitary transformations scales linearly with the number of states $N$~\cite{complexity}, these findings suggest that the unitary transformation landscape appears to be structurally similar to the ensemble control landscape with fully nondegenerate $\rho$ and $O$ matrices. \\
\indent Figure \ref{disthistW} displays the histograms resulting from calculating the pairwise distances between the control fields using Eq.~(\ref{pairwise}), as explained in Section~5.1.1. The two sets of control trajectories, each of size 250, were used to produce the histograms in Figure~\ref{disthistW}(a) and \ref{disthistW}(b) which satisfy $R<1.34$ and $R>1.48$, respectively. Significantly, the histograms computed with the lowest values of $R$ and with the highest values of $R$ are almost identical, indicating that trajectories with varying values of $R$ for generating unitary transformations are distributed in the same way across control space, as found in Figure~\ref{disthistpO} for the ensemble control landscape. This result was also found for a five level state-to-state transition probability landscape~\cite{Nanduri}, suggesting that similar distributions of fields for high and low $R$ trajectories is a general property of quantum control landscapes. However, the unitary transformation landscape explored here additionally possesses similar distance distributions between all pairs of initial fields, final fields, and initial-final fields. This additional feature is not shared by the quantum ensemble and state-to-state transition probability landscapes.

\subsection{Interactions of optimization trajectories with saddle submanifolds}
\indent Although the saddle submanifolds of the quantum ensemble and unitary transformation landscapes are not able to prevent reaching the global optimum, they may still significantly influence the optimization trajectories. In order to examine the extent to which saddle submanifolds affect the climb up the landscape, the metrics set out in Section 3.1.3 and 3.2.3 (and Appendices A and B, respectively) were used to calculate the distance from each point along a landscape climb trajectory to all critical submanifolds. As examples, the quantum ensemble control landscape was generated by the initial density matrix $\rho_2$ and observable operator $O_1$ from Eqs.~(\ref{rhos12}) and (\ref{Os12}), respectively, while the unitary control landscape was generated with $W$ in Eq.~(\ref{W}).

\subsubsection{Quantum ensemble control landscape}
\begin{figure}[h!]
  \centering
  \includegraphics[width=\textwidth]{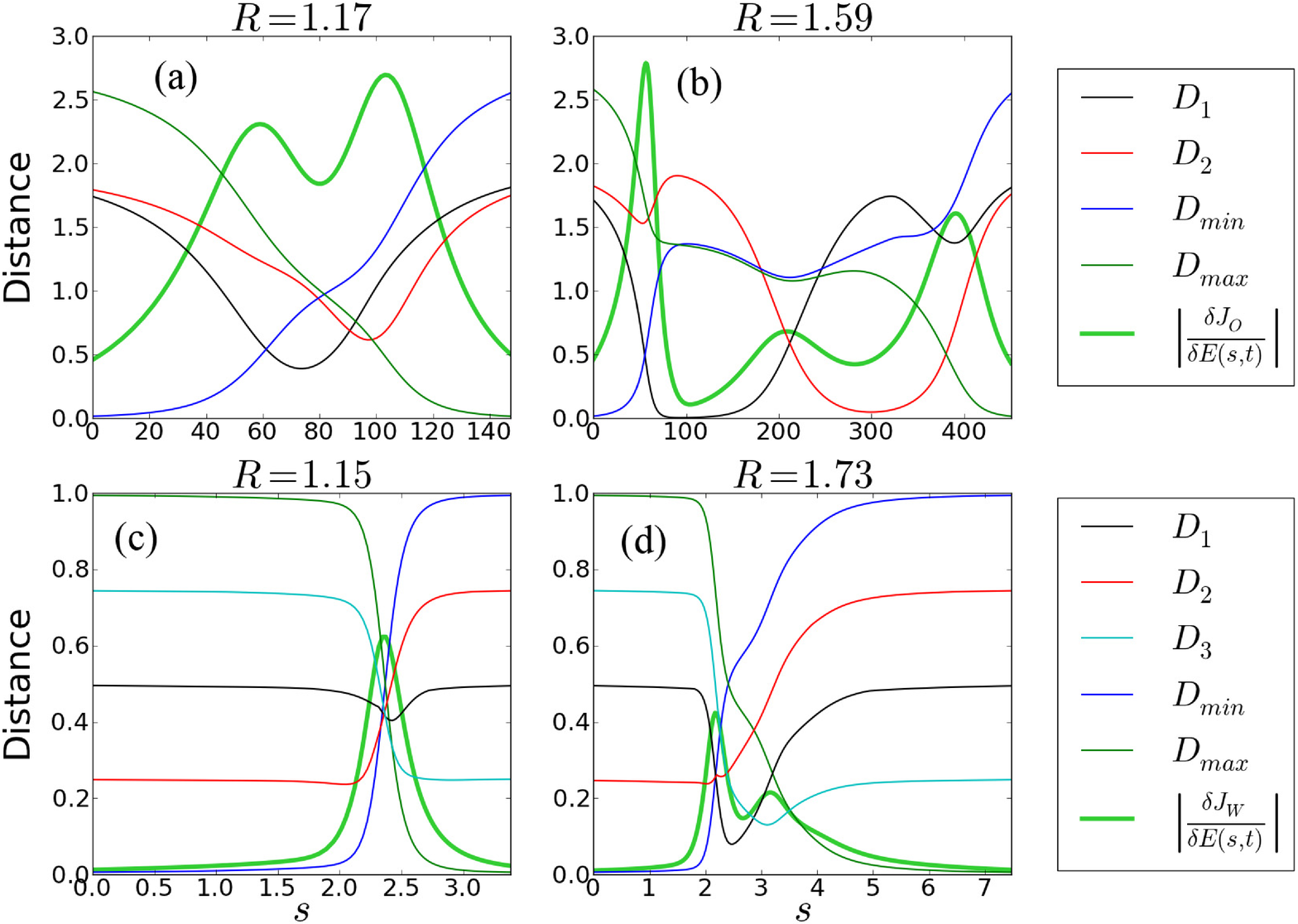}
  \caption{In (a) and (b), the distances $D$ to the critical submanifolds are plotted using Eq.~(\ref{saddlemetricTrpO}) along a low and high $R$ value optimization trajectory, respectively, on a quantum ensemble control landscape. It can be seen in (b) that the distance to saddle submanifolds 1 and 2 become very small along the high $R$ trajectory, indicating that the saddles distorted the trajectory by acting as attractors. The norm of the gradient function along the climb, shown as the thick green curve, also noticeably dips in (b), further confirming that saddle submanifolds were encountered. For a unitary transformation control landscape in (c) and (d), the distances $D$ to the critical submanifolds, using Eq.~(\ref{saddlemetricW}), are plotted along a low and high $R$ value optimization trajectory. In (c), no saddle submanifolds are encountered, but in (d) the black and green curves (i.e., for saddles 1 and 3, respectively) dip sharply along with the norm of the gradient, signaling attraction towards these saddle submanifolds and resulting in the corresponding high value of $R$.}
  \label{splitsads}
\end{figure}

\indent The values of the critical submanifolds of the system on the quantum ensemble control landscape~\cite{BigTrpO} of $\rho_2$ and $O_1$ are given by $J_O^{\mathrm{min}}=0$, $J_O^1=\frac{1}{9}$, $J_O^2=\frac{5}{36}$, and $J_O^{\mathrm{max}}=\frac{1}{4}$. Here, $J_O^1$ and $J_O^2$ correspond to saddles, although a trajectory passing through these values may not actually come close to the associated submanifold of saddle controls. The normalized distances $D$ to these four critical submanifolds, using Eq.~(\ref{saddlemetricTrpO}), are computed during the optimizations with the lowest and highest $R$ values out of the 1000 runs performed, which were $R=1.17$ and $R=1.59$, respectively. The results are shown in Figures~\ref{splitsads}(a) and \ref{splitsads}(b). The magnitude of the gradient for each trajectory is also plotted to show the rate of climbing, as an optimization typically slows down near a critical submanifold. In Figure~\ref{splitsads}(a), the run with a small $R$ value does not closely approach any of the saddle submanifolds. This observation is corroborated by examining the magnitude of the gradient, which falls only slightly in the middle of the climb. The gradient is naturally small near the beginning and end of the climb, where the trajectory leaves (approaches) the critical submanifolds corresponding to the bottom (top) of the landscape. In contrast, Figure~\ref{splitsads}(b) shows that the run with a high $R$ value hovers in the vicinity of the two saddle submanifolds for extended periods during the climb, indicated by the distance to the first and second saddle manifolds, $D_1$ and $D_2$, respectively, becoming very small during portions of the run. Concurrently, the magnitude of the gradient falls sharply in those two areas, recovering only when $D_2$ starts to rise, thus confirming that the run passed closely by these two saddle submanifolds. Comparison of Figures~\ref{splitsads}(a) and \ref{splitsads}(b) suggests that control trajectories that are `diverted' to pass closely by saddle submanifolds can have higher $R$ values.

\subsubsection{Unitary transformation control landscape}
\indent Figures~\ref{splitsads}(c) and \ref{splitsads}(d) show the distances to the critical submanifolds while traversing the unitary transformation control landscape using Eq.~(\ref{saddlemetricW}); the magnitude of the gradient is also shown. Figure~\ref{splitsads}(c) corresponds to a run with a low $R$ value of 1.15, and Figure~\ref{splitsads}(d) corresponds to a run with a high $R$ value of 1.73. The distances to the saddle submanifolds $D_1$ and $D_3$ assume lower values during the high $R$ run (as compared to the low $R$ run), and correspondingly the magnitude of the gradient also dips when the distances to $D_1$ and $D_3$ are at a minimum. Neither of the trajectories approach the second saddle submanifold, as indicated by the behavior of $D_2$ on the trajectory. These plots also support the notion that saddle submanifolds are capable of attracting nearby paths of steepest descent away from proceeding in a direct manner to a global minimum.

\subsection{Hessian-gradient eigen-relation}
\begin{figure}[h!]
  \centering
  \includegraphics[width=\textwidth]{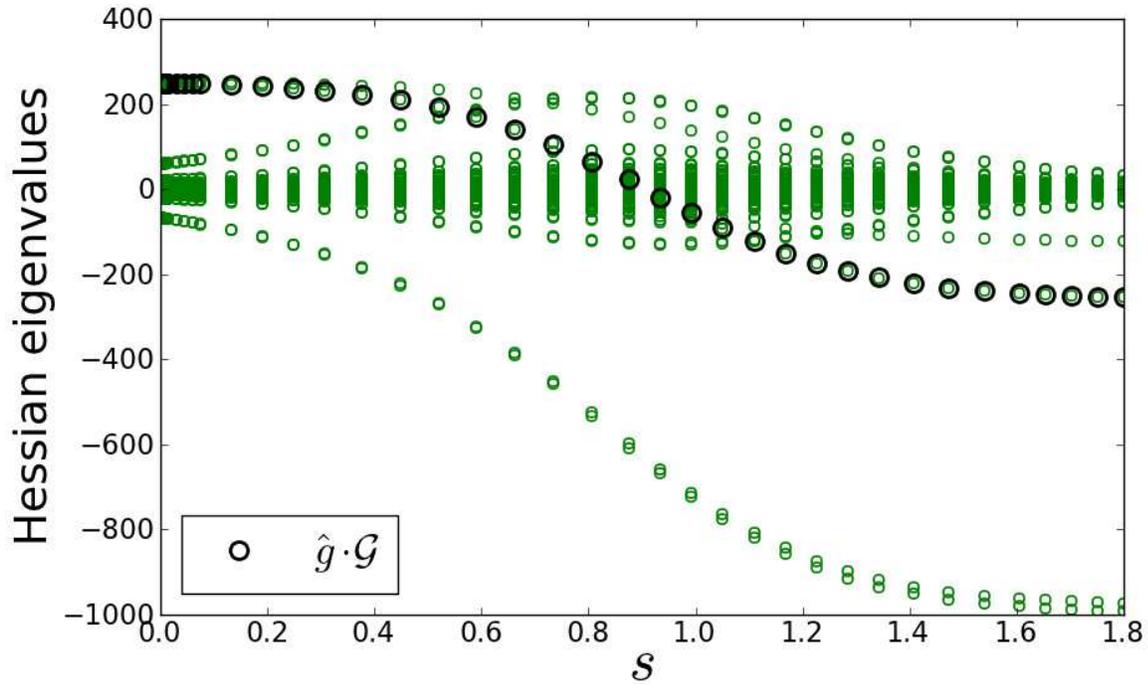}
  \caption{The thick black circles are values of $\hat{g}(s)\cdot \mathcal{G}(s)=\hat{g}(s)\cdot \mathcal{H}(s)\cdot \hat{g}(s)$ along a nearly straight shot to the top of a state-to-state transition probability landscape. Here $\mathcal{H}(s)$ is the Hessian and $\hat{g}$ is the normalized gradient function. The eigenvalues of the Hessian are shown as the smaller green circles, for each point $s$ along the climb of the landscape. The black circles match with a green circle at each point on the climb, supporting the identification of the gradient with an eigenfunction of the Hessian. As the associated eigenvalue of the Hessian changes from positive to negative, the function $\mathcal{G}(s)=\mathcal{H}(s)\cdot \hat{g}$ changes from being parallel to antiparallel to the gradient function.}
  \label{dot}
\end{figure}

\begin{figure}[h!]
  \centering
  \includegraphics[width=.8\textwidth]{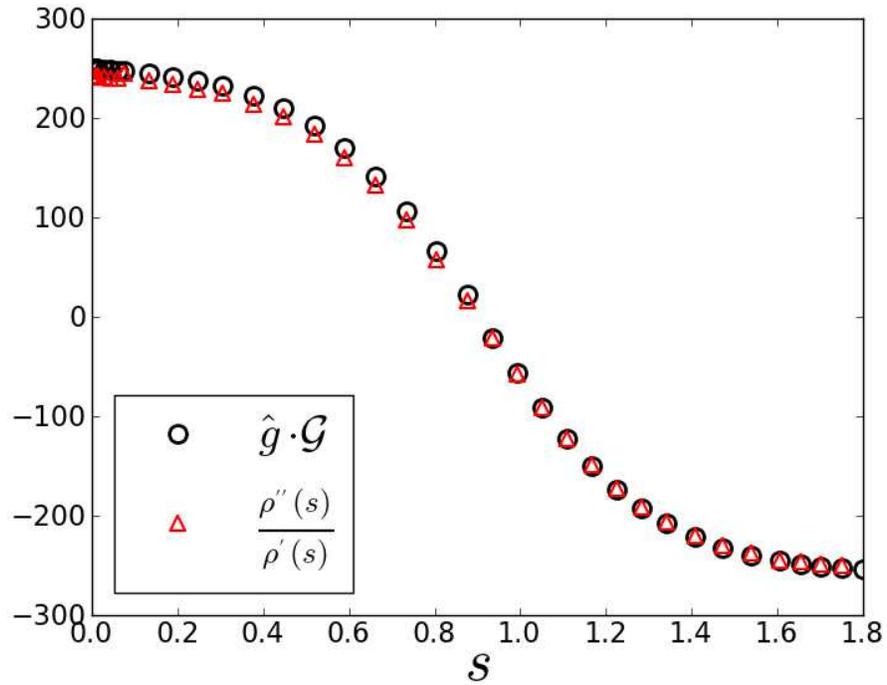}
  \caption{A plot of $\hat{g}\cdot \mathcal{G}=\hat{g}\cdot \mathcal{H}\cdot \hat{g}$, which is the factor by which the Hessian $\mathcal{H}$ scales the normalized gradient $\hat{g}$. From Eq.~(\ref{eigenrelation}), along a straight control trajectory we should have $\hat{g}\cdot \mathcal{G}=\frac{\rho''(s)}{\rho'(s)}$. Thus, the plot also shows $\rho''(s)/\rho'(s)$ evaluated along the gradient landscape climb. The good agreement between the two functions signifies the validity of Eq.~(\ref{eigenrelation}) along a straight control trajectory.}
  \label{match}
\end{figure}

\indent The results of Section 4 identify that the gradient of the cost functional should be an eigenfunction of the Hessian along a straight trajectory. Here, we provide evidence to verify the Hessian-gradient eigen-relation for nearly straight shots found in simulations. In order to illustrate this phenomenon as clearly as possible, we consider a simple three level quantum system with Hamiltonian and dipole matrix
\begin{equation}
H_0=
\begin{pmatrix}
-10 & 0 & 0 \\ 0 & -5 & 0 \\ 0 & 0 & 5
\end{pmatrix}
\phantom{space}
\mu=
\begin{pmatrix}
0 & -1 & -0.5 \\ -1 & 0 & 1 \\ -0.5 & 1 & 0
\end{pmatrix},
\end{equation}
and we use the cost functional of Eq.~(\ref{TrpO}), with
\begin{equation}
\rho=
\begin{pmatrix}
1 & 0 & 0 \\ 0 & 0 & 0 \\ 0 & 0 & 0
\end{pmatrix}
\phantom{space}
O=
\begin{pmatrix}
0 & 0 & 0 \\ 0 & 0 & 0 \\ 0 & 0 & 1
\end{pmatrix},
\end{equation}
corresponding to maximizing the pure state transition probability $P_{1\to 3}(T)$ with $T=10$. This landscape contains no saddle submanifolds, and therefore the control trajectories tend to have smaller values of $R$. We used the stochastic Particle Swarm Optimization (PSO) algorithm~\cite{PSO,Kennedy-PSO} to locate an initial control $E(0,t)$ taking the form in Eq.~(\ref{parametrized}) with $M=20$ that produced a nearly straight trajectory with $R=1.002$. In order to assess whether a lower bound for $R$ (other than 1.0) might exist, a Derandomized Evolution Strategy (ES)~\cite{Hansen01completely,HansenDR2PPSN08} was used in a more extensive search for a straight path. Using this algorithm together with a high-order Gaussian quadrature rule to accurately evaluate the integrals in Eqs.~(\ref{pathlength}) and (\ref{Euclength}) allowed reaching an even lower value of $R=1.00003$. As this result was at the limit of our numerical accuracy, we cannot rule out the prospect of finding a still lower value of $R$. Below, the eigen-relation is tested with the former conservative near straight trajectory of $R=1.002$. \\
\indent We seek to affirm Eq.~(\ref{eigenrelation}), which states that the gradient function aligns with some eigenfunction of the Hessian. As in Eq.~(\ref{eigenrelation}), we simplify notation with the Hessian and gradient given by $\mathcal{H}(s)$ and $g(s)$, respectively, and a `dot' $\cdot$ denotes integration over time $t\in[0,T]$. Thus, we calculated $\hat{g}(s)\cdot\mathcal{G}(s)$, where $\mathcal{G}(s)=\mathcal{H}(s)\cdot \hat{g}(s)$ is the function resulting from the action of the Hessian on the unit gradient function, $\hat{g}(s)=\frac{\delta P_{1\to 3}}{\delta E(s,t)}/\left(\frac{\delta P_{1\to 3}}{\delta E(s,t)}\cdot \frac{\delta P_{1\to 3}}{\delta E(s,t)}\right)^{\frac{1}{2}}$. In Figure~\ref{dot}, $\hat{g}\cdot \mathcal{G}(s)=\hat{g}\cdot \mathcal{H}\cdot \hat{g}$ is plotted versus $s$ as the thick black circles, along with every eigenvalue of the Hessian shown as the smaller green circles. Intriguingly, at every point on the climb, the black circle overlaps almost exactly with one of the green circles, suggesting that the gradient function can be identified with some eigenfunction of the Hessian. The sign of $\hat{g}(s)\cdot\mathcal{G}(s)$ reveals that $\mathcal{G}$ and $g$ are parallel at the beginning of the climb and antiparallel at the end, which is consistent with the assessment discussed below Eq.~(\ref{criticaleigenvector}) \\
\indent We will now illustrate the validity of Eq.~(\ref{eigenrelation}), rewritten in the notation here as $\hat{g}\cdot \mathcal{H}\cdot \hat{g}\equiv \hat{g}\cdot \mathcal{G}=\rho''(s)/\rho'(s)$. This assessment is performed by replotting $\hat{g}(s)\cdot\mathcal{G}(s)$ versus $s$ in Figure~\ref{match} along with $\rho''(s)/\rho'(s)$, calculated for this control trajectory. According to Eq.~(\ref{straightshotderivative}), for a $R=1$ straight control trajectory, $\frac{\delta P_{1\to 3}}{\delta E(s,t)}=\rho'(s)\; \Delta E(t)$. Since the trajectory considered here is not exactly straight with $R=1.002$, we determined the average value of $\rho'(s)$ by setting $\rho'(s) = \frac{1}{T}\int_0^T \frac{\delta P_{1\to 3}}{\delta E(s,t)}\frac{1}{\Delta E(t)}dt$, where $\Delta E(t)=E(s_{max},t)-E(0,t)$ and $\rho''(s)$ was then obtained by numerically differentiating $\rho'(s)$. Examining Figure~\ref{match} reveals that $\hat{g}\cdot \mathcal{G}\simeq \rho''(s)/\rho'(s)$ to good accuracy, providing evidence for the validity of the eigen-relation derived in Section 4.

\section{Conclusions}
\indent The exploration of landscape structure first performed for the state-to-state transition probability~\cite{Nanduri} is extended in this paper to more general quantum ensemble and unitary transformation control landscapes. These collective results point towards the existence of simple quantum control landscape structure, which was first encountered experimentally~\cite{roslund} as well as hinted at by preliminary numerical studies~\cite{complexity,hamiltonianstructure}. \\
\indent A special feature of the landscapes considered in this work is the presence of critical submanifolds that possess saddle topology. These submanifolds do not prevent optimization trajectories from reaching the top of the landscape, but they can have an impact on the search effort, reflected in the identified $R$ value. We find that the presence of saddle submanifolds on the landscape has a small but discernible effect on the ease of optimization reflected in the distortion of the control trajectories. Importantly, the appearance of saddle submanifolds did not cause $R$ to take on values in excess of 2.0 for the systems examined here, indicating that these quantum control landscapes, in addition to state-to-state transition probability landscapes~\cite{Nanduri}, possess markedly simple structure. \\
\indent A condition for the existence of control trajectories with $R=1.0$, which would indicate exceedingly simple landscape structure, was also derived. It was found that in order for the gradient algorithm to yield a straight trajectory, the gradient function must not only be separable in $s$ and $t$, but must also be an eigenfunction of the Hessian. This situation also implies that all higher-order derivatives of the cost functional possess a common eigenfunction. The eigen-relation may form a basis to identify the conditions under which exactly straight control trajectories derived from the gradient algorithm exist. Additional simulations exploring landscape structure on a wider variety of landscapes would also allow for a better understanding of the conditions that lead to low values of $R$. One direction for further investigation is whether achieving control of unitary transformations, which requires control over every matrix element in the propagator, leads to higher values of $R$ than even for the fully nondegenerate ensemble control landscapes. A simple assessment of this matter was performed for a two level system using an ES algorithm~\cite{Hansen01completely,HansenDR2PPSN08} to search for straight trajectories. For two-level systems, in the cases of maximizing $P_{1\to 2}$ we find $R-1\leq 10^{-6}$, while for unitary transformation control $R=1.0016$ was achieved. \\ 
\indent Quantum control landscapes are expected to nominally be very complex, as they are defined on extremely high dimensional spaces and represent the generally nonlinear relationship, encoded through Schr\"{o}dinger's equation, between the unitary propagator and a dynamical set of controls. Thus, a natural expectation is that trajectories traveling over these landscapes would require considerable effort to reach an optimal outcome. However, the topology of quantum control landscapes is very simple when certain reasonable physical assumptions are met, providing one factor to explain the relative ease of performing optimizations~\cite{Moorewithcontrolstuff,surveypaper,topologyscience,kosut}. These topological results are complemented by the finding in this work of simple landscape structure. This study, along with Ref.~\cite{Nanduri}, supports the notion that simple structure is a universal and robust property of quantum control landscapes. Collectively, the twofold simplicity of landscape features, encapsulated by topology and structure, provides a foundation for more clearly explaining the widespread efficiency of finding optimal controls over quantum phenomena.

\newpage
\appendixtitleon
\begin{appendices}

\section{Critical submanifolds of the quantum ensemble control landscape}\label{sadpO}

This Appendix gives a brief overview of the critical submanifolds of the quantum ensemble control landscape, along with a metric which can be used to determine the distance to all critical submanifolds from any given point in control space. A complete characterization of the critical submanifolds is given in Ref.~\cite{Rebing}. \\
\indent The analysis of Section 3.1 reveals that at a critical point, $[O,\rho(T)]=0$. This condition implies that the number of critical submanifolds is in one-to-one correspondence with the number of ways the eigenvalues of $\rho=\rho(0)$ can be permuted such that their overlaps with the eigenvalues of $O$ are distinct. These permutations are induced by critical propagators $U(T,0)$, each of which belongs to a distinct critical submanifold, and can be enumerated using contingency tables~\cite{Rebing}. These tables possess $n$ rows and $m$ columns and can be constructed as follows. First, $\rho$ and $O$ are expressed in diagonal form with their eigenvalues in descending order. Then, a particular permutation is applied to the eigenvalues of $\rho$; such a permutation matrix $U_o$ is associated with the said critical submanifold. The number of times the eigenvalue $\lambda_i$ of $\rho$ overlaps with the eigenvalue $\epsilon_j$ of $O$, after the permutation is applied, is the value $c_{ij}$ of the $(i,j)$ entry in the contingency table. The rows of the contingency table must add up to the sum of the degeneracies of the eigenvalues of $O$, and the columns of the table must add up to the sum of the degeneracies of the eigenvalues of $\rho$. This fact can be used to easily construct all possible contingency tables and therefore reveal every critical submanifold along with its associated height on the landscape. \\
\indent The contingency tables can also be used to calculate a distance to a particular critical submanifold from any point on the landscape. The distance is expressed in terms of the unitary transformation $U[E(t)]$ corresponding to a particular control field $E(t)$. A short exposition of this metric is given here; a full discussion is given in Ref.~\cite{distancemetrics}. Consider a point on the landscape generated by $U_o$ belonging to some critical submanifold; i.e, $U_o$ is a permutation matrix. The matrix $U_o$ may be divided into blocks $\{U_o^{ij}\}$, starting in the upper left corner, of sizes $\{o_i \times p_j\}$, where $o_i$ is the degeneracy of the $i$th eigenvalue of $O$, and $p_j$ is the degeneracy of the $j$th eigenvalue of $\rho(0)$, with the eigenvalues arranged in descending order. It can be shown that the singular value decomposition of each block contains only 0s and 1s, and that for the $ij$th block of size $o_i\times p_j$ the number of 1s is equal to $c_{ij}$, which is the entry in the $i$th row and $j$th column in the contingency table. This fact can be used to specify the distance to each critical submanifold. For each block and its associated entry $c_{ij}$, define an $N\times N$ diagonal matrix $S_{ij}$, which has $c_{ij}$ values of 1 starting in the upper left corner in the diagonal:
\begin{equation}
S_{ij}=\begin{pmatrix} \mathbbm{I}_{c_{ij}} &  \\  & 0
\end{pmatrix}.
\end{equation}
To find the distance of an arbitrary unitary matrix $U$ from the critical submanifold $U_o$, we divide $U$ into blocks $\{U^{ij}\}$ in the manner described above. Each block $U^{ij}$ of $U$ is written in its singular form $\Sigma_{ij}$, with the singular values on the diagonal sorted in descending order. For each block in $U$, we define $V_{ij}=S_{ij}-\Sigma_{ij}$. The distance from $U$ to the critical submanifold of $U_o$ is then
\begin{equation}\label{saddlemetricTrpO}
D=\sum_{i=1}^{p}\sum_{j=1}^o\tr \left (V_{ij}^{\dagger}V_{ij}\right ),
\end{equation}
where $p$ and $o$ are the number of distinct eigenvalues of $\rho(0)$ and $O$, respectively. It is convenient to normalize this metric~\cite{SVDnormalization}. To do so, the unitary matrix $U$ can be assigned a vector consisting of the singular values of all of its blocks $\{U^{ij}\}$. Since the squares of these singular values sum to $N$, each vector can be thought of as lying in the spherical polygon spanned by the vectors corresponding to the $U_o$ matrices of the critical submanifolds. This formulation implies that the point which is the greatest distance from a critical submanifold is another vertex, which itself corresponds to another critical submanifold. By calculating the pairwise distances between all the critical submanifolds, we can find the maximum distance any point can be from a particular critical submanifold, and then normalize the metric by dividing all distances to that critical submanifold by the maximum distance, thereby giving $0\leq D \leq 1$. This normalization is used in the results reported in Figures~\ref{splitsads}(a) and \ref{splitsads}(b).

\section{Critical submanifolds of the unitary transformation control landscape}\label{sadW}
\indent Here we outline a distance metric for the unitary transformation control landscape; full details are given in Ref.~\cite{distancemetrics}. Recall, from Section 3.2 that for a unitary transformation $U_o$ corresponding to a critical point, the eigenvalues of $W^{\dagger}U_o$ are $\pm 1$. The distance from an arbitrary unitary matrix $U$ to a submanifold containing the critical point $U_o$ can be specified by noting that the eigenvalues of $W^{\dagger}U$ take values on the unit circle in the complex plane. The $i$th eigenvalue of $W^{\dagger}U$ is $\lambda_i$, and the eigenvalues are sorted in ascending order by their real parts. To find the distance to the critical submanifold which corresponds to $\alpha$ eigenvalues of -1 and $N-\alpha$ eigenvalues of 1 for $W^{\dagger}U_o$, we compare the real part of the smaller $\alpha$ eigenvalues of $W^{\dagger}U$ with -1, and the real part of the larger $N-\alpha$ eigenvalues with 1~\cite{distancemetrics}. The distance between $U$ and the critical submanifold containing $U_o$ then is
\begin{equation}\label{saddlemetricW}
D=\left[\sum_{i=0}^{\alpha} \left (1+\mbox{Re} \{\lambda_i\} \right ) + \sum_{\alpha}^{N-1} \left (1-\mbox{Re} \{\lambda_i\} \right )\right]/2N.
\end{equation}
The maximum distance to a critical submanifold in this case is $2N$, and the denominator of Eq.~(\ref{saddlemetricW}) assures that $0\leq D\leq 1$.

\end{appendices}

\section*{Acknowledgments}
We thank Greg Riviello and Julia Yan for helpful discussions. A.N. acknowledges support from the Program in Plasma Science and Technology at Princeton University. We also acknowledge support from ARO Grant No. W911NF-13-1-0237, ARO-MURI Grant No. W911NF-11-1-2068, DOE Grant No. DE-FG02-02ER15344, and NSF Grant No. CHE-1058644.


\begin{thebibliography}{99}

\bibitem{balint-kurti}
G. G. Balint-Kurti, S. Zou, and A. Brown, \textit{Advances in Chemical Physics}, page 43. Wiley, 2008.

\bibitem{HDR}
T.-S. Ho, J. Dominy, and H. Rabitz, Phys. Rev. A \textbf{79}, 013422 (2009).

\bibitem{Dong}
D. Dong and I. R. Peterson, IET Control Theory Appl. \textbf{4}, 2651 (2010).

\bibitem{Moorewithcontrolstuff}
K. W. Moore and H. Rabitz, Phys. Rev. A \textbf{84}, 012109 (2011).

\bibitem{W}
K. W. M. Tibbetts, C. Brif, M. D. Grace, A. Donovan, D. L. Hocker, T.-S. Ho, R. B. Wu, and H. Rabitz, Phys. Rev. A \textbf{86}, 062309 (2012).

\bibitem{Blasi}
T. Blasi, M. F. Borunda, E. R\"{a}s\"{a}nen, and E. J. Heller, Phys. Rev. B \textbf{87} 241303 (2013).

\bibitem{Nest}
M. Nest, M. Ludwig, I. Ulusoy, T. Klamroth, and P. Saalfrank, J. Chem. Phys. \textbf{138}, 164108 (2013).

\bibitem{Doria}
P. Doria, T. Calarco, and S. Montangero, Phys. Rev. Lett. \textbf{106}, 190501 (2011).

\bibitem{Rahmani}
A. Rahmani, T. Kitagawa, E. Demler, and C. Chamon, Phys. Rev. A \textbf{87}, 043607 (2013).

\bibitem{Rasanen}
E. R\"{a}s\"{a}nen and E. J. Heller, Eur. Phys. J. B \textbf{86}, 1 (2013).

\bibitem{Young}
C. Brif, M. D. Grace, M. Sarovar, and K. C. Young, arXiv:1310:3443v1.

\bibitem{lasers}
R. S. Judson and H. Rabitz, Phys. Rev. Lett. \textbf{68}, 1500 (1992).

\bibitem{pulseshaping}
A. Weiner, Rev. Sci. Instrumen. \textbf{71}, 1929 (2000).

\bibitem{Bucker}
R. B\"{u}cker, T. Berrada, S. van Frank, J. Schaff, T. Schumm, J. Schmiedmayer, G. J\"{a}ger, J. Grond, and U. Hohenester, J. Phys. B \textbf{46}, 104012 (2013).

\bibitem{Truong}
N. X. Truong, P. Hilse, S. G\"{o}de, A. Przystawik, T. D\"{o}ppner, Th. Fennel, Th. Bornath, J. Tiggesb\"{a}umer, M. Schlanges, G. Gerber, and K. H. Meiwes-Broer, Phys. Rev. A \textbf{81}, 013201 (2010).

\bibitem{Hildner}
R. Hildner, D. Brinks, J. B. Nieder, R. J. Cogdell, and N. F. van Hulst, Science \textbf{340}, 1448 (2013).

\bibitem{Nanduri}
A. Nanduri, A. Donovan, T.-S. Ho, and H. Rabitz, Phys. Rev. A \textbf{88}, 033425 (2013).

\bibitem{surveypaper}
R. Chakrabarti and H. Rabitz, Int. Rev. Phys. Chem. \textbf{26} 671 (2007).

\bibitem{Wsaddlepoints}
H. Rabitz, M. Hsieh, and C. Rosenthal, Phys. Rev. A \textbf{72}, 052337 (2005).

\bibitem{Wsaddlepoints2}
M. Hsieh and H. Rabitz, Phys. Rev. A \textbf{77}, 042306 (2006).

\bibitem{Wstructure}
K. W. Moore, R. Chakrabarti, G. Riviello, and H. Rabitz, Phys. Rev. A \textbf{83}, 012326 (2011).

\bibitem{BigTrpO}
T.-S. Ho and H. Rabitz, J. Photochem. Photobiol. A \textbf{180}, 226 (2006).

\bibitem{Rebing}
R. Wu, H. Rabitz, and M. Hsieh, J. Phys. A, \textbf{41}, 015006 (2008).

\bibitem{singularities}
R.-B. Wu, R. Long, J. Dominy, T.-S. Ho, and H. Rabitz, Phys. Rev. A \textbf{86}, 013405 (2012).

\bibitem{TrpOtopology}
M. Hsieh, R. Wu, and H. Rabitz, J. Chem. Phys. \textbf{130}, 104109 (2009).

\bibitem{roslund}
J. Roslund and H. Rabitz, Phys. Rev. A \textbf{80}, 013408 (2009).

\bibitem{complexity}
K. W. Moore, M. Hsieh, and H. Rabitz, J. Chem Phys. \textbf{128}, 154117 (2008).

\bibitem{hamiltonianstructure}
A. Donovan, V. Beltrani, and H. Rabitz, Phys. Chem. Chem. Phys. \textbf{13}, 7348 (2011).

\bibitem{Nielsen}
M. A. Nielsen and I. L. Chuang, \textit{Quantum Computation and Quantum Information}, Cambridge University Press, 2004.

\bibitem{DMORPH}
A. Rothman, T.-S. Ho, and H. Rabitz, Phys. Rev. A \textbf{72}, 023416 (2005).

\bibitem{otherDMORPH}
A. Rothman, T.-S. Ho, and H. Rabitz, J. Chem. Phys. \textbf{123}, 134104 (2005).

\bibitem{eig1}
R. Magnanini, Appl. Anal. \textbf{41}, 171 (1991).

\bibitem{eig2}
D. Drucker and S. A. Williams, Am. Math. Mon. \textbf{110}, 869 (2003).

\bibitem{PSO}
J. Kennedy and R. Eberhart, in Proceedings of the IEEE International Conference on Neural Networks, Perth, WA, 1995 (IEEE, Piscataway, NJ, 1995), Vol. 4, p. 1942.

\bibitem{Kennedy-PSO}
J. Kennedy and R. Eberhart, \textit{Swarm Intelligence} (Morgan Kaufman, San Francisco, CA, 2001).

\bibitem{Hansen01completely}
N. Hansen and A. Ostermeier, Evol. Comput. \textbf{9}, 159 (2001).

\bibitem{HansenDR2PPSN08}
R. Ros and N. Hansen, Lect. Notes Comput. Sc. \textbf{5199}, 296 (2008).

\bibitem{topologyscience}
H. Rabitz, M. Hsieh, and C. Rosenthal, Science \textbf{303}, 1998 (2004).

\bibitem{kosut}
H. Rabitz, T.-S. Ho, M. Hsieh, R. Kosut, and M. Demiralp, Phys. Rev. A \textbf{74}, 012721 (2006).

\bibitem{distancemetrics}
Q. Sun, R. Wu, G. Riviello, and H. Rabitz (to be published).

\bibitem{SVDnormalization}
G. Riviello and H. Rabitz, (to be published).

\end{thebibliography}

\newpage
\section*{Figures}

\newpage

\newpage


\end{document}